\documentclass[12pt, preprint]{aastex}
\usepackage{graphicx}  
\usepackage{dcolumn}   
\usepackage{bm}        
\usepackage{amssymb}   
\usepackage{color}
\usepackage{epsfig}
\usepackage{amsmath}

\def\apj{Astrophys. J.}
\def\apjl{Astrophys. J. Lett.}
\def\apjs{Astrophys. J. Supp. Ser. }

\def\aap{Astron. Astrophys. }

\def\mnras{Mon. Not. Roy. Astron. Soc. }

\def\nat{Nature}

\def\icarus{Icarus}

\def\be{\begin{equation}}
\def\ee{\end{equation}}
\def\ba{\begin{eqnarray}}
\def\ea{\end{eqnarray}}
\def\go{\mathrel{\raise.3ex\hbox{$>$}\mkern-14mu
             \lower0.6ex\hbox{$\sim$}}}
\def\lo{\mathrel{\raise.3ex\hbox{$<$}\mkern-14mu
             \lower0.6ex\hbox{$\sim$}}}

\def\tomega{{\tilde{\omega}}}

\begin{document}

\title{Protoplanetary Disk Resonances and Type I Migration}
\author{David Tsang} \affil{TAPIR, California Institute of Technology, Pasadena, CA, USA; dtsang@caltech.edu}

\date{\today}

\begin{abstract}
Waves reflected by the inner edge of a protoplanetary disk are shown to significantly modify Type I migration, even allowing the trapping of planets near the inner disk edge for small planets in a range of disk parameters. This may inform the distribution planets close to their central stars, as observed recently by the Kepler mission.
\end{abstract}
\maketitle

\section{Introduction}
The location and configuration of planets within their planetary systems depend sensitively on the effects  of many physical processes including planet-planet scattering \citep[e.g.][]{Rasio96}, Kozai oscillations \citep[e.g.][]{Fabrycky07}, and disk migration \citep[e.g.][hereafter GT80]{GT80}\footnote{For review and comparison of migration mechanisms see \citet{Morton11} and \citet{Ward00}, and references therein.}. Of these, disk migration is likely to have been dominant in systems where planets are observed to be co-planar \citep[e.g.][]{Steffen10}. 
The density waves launched inwards from the inner Lindblad resonances carry away negative angular momentum driving the planet outwards (positive torque). The waves launched outwards from the outer Lindblad resonances carry away positive angular momentum driving it inwards  (negative torque). 
For planets too small to clear a gap in their protoplanetary disk  Type I migration occurs due to the asymmetry between the Lindblad torques \citep[hereafter W86 and W97]{Ward86, Ward97} with the outer torque being typically stronger, driving a net migration inwards. 

Recent observations by the Kepler spacecraft have shed new light on to the distribution of extrasolar planets within their planetary systems. \citet{Howard11} examined the radial distribution of planets closer than 0.25 AU of solar type stars observed in the Kepler field. They found that smaller planets ($R_p \leq 2-4 R_{\earth}$) have an exponential cutoff in the radial distribution at an orbital period $ \sim 7$ days, while larger planets have exponential cutoffs closer to the star at $\sim 2$ days.

The protoplanetary disks are likely be truncated before direct contact with the host stars by a strong stellar magnetic field  near to the corotation radius \citep[see e.g.][]{Shu94}. Rotational studies of classical T Tauri stars give periods ranging from $\sim 3-7$ days for stars with $M_* > 0.25 M_\odot$ \citep{Lamm05}. The truncation radius is typically taken to be $\sim 5$ stellar radii, or about $ \sim 4$ day periods for typical 2 solar radius pre-main-sequence stars \citep{Gullbring98}. 
The inner edge of the accretion disk plays a role in halting migration very close to the host star; disk migration is naturally halted once a planet passes through the inner disk edge \citep{Lin96} and the Lindblad resonances can no longer interact with the disk material. 

Close to the inner edge of the disk the effect of the corotation resonance also becomes important and can exceed the differential Lindblad torque, particularly at locations with strong gradients in the surface density profile \citep{GT79}. However, unless the viscosity is strong enough to transfer angular momentum away from the co-orbital region quickly enough, the corotation resonance saturates and the corotation torque goes to zero \citep{Masset01, Ogilvie03, Baruteau08}. 

In this letter we will outline a halting mechanism for Type I migration of planets due to density waves resonant between the inner Lindblad resonance and the disk inner edge. This allows for the Type I migration of smaller planets to be halted within the disk, at a few times the inner disk radius.

\section{Torque due to Lindblad Resonances}
\subsection{Basic Equations}
To evaluate the effect of a perturbing body on a gas disk we limit ourselves to the examination of the 2-dimensional disk and proceed with the vertically integrated disk variables $\Sigma \equiv \int \rho~dz$ and $P \equiv \int p ~dz$ the surface density and vertically integrated pressure respectively. 
We assume that in the unperturbed state the disk velocity profile is Keplerian, ${\bf u}_o = r \Omega \hat{\phi}$ and that perturbations are barotropic such that $P = P(\Sigma)$.

We begin by first considering the inviscid limit of the system.
The linear perturbation equations for continuity and momentum conservation are
\be
\frac{\partial \delta \Sigma}{\partial t} + \nabla \cdot ( \Sigma \delta {\bf u} + {\bf u}_o \delta \Sigma) = 0\, , 
\ee
\be
\frac{\partial \delta {\bf u}}{\partial t} + ({\bf u}_o \cdot \nabla) \delta {\bf u} + (\delta {\bf u} \cdot \nabla) {\bf u}_o) = -\frac{1}{\Sigma} \nabla \delta P + \frac{\delta \Sigma}{\Sigma^2} \nabla P - \nabla \Phi,
\ee
where $\delta$ signifies the linear perturbation of a quantity, $\Phi$ is the external potential due to the perturbing planet with mass $M_p$ located at radius $r_p$ away from the central star.

For a perturber with vanishing eccentricity we can break down this perturbing potential into azimuthal Fourier components \citep[hereafter GT79]{GT79}
\be
\Phi = \sum_{m=0}^\infty \Phi_m (r) \exp(im\phi - im\Omega_p t)\, ,
\ee
where $\Omega_p$ is the orbital frequency of the perturber, and 
\be
\Phi_m \equiv -\frac{GM_p}{2r_p} (2-\delta_{m,0}) \left(b_{1/2}^m(r/r_p) - \frac{\Omega_p^2 r_p^2 r}{G M_p}\delta_{m,1}\right)\,,
\ee
$\delta_{i,j}$ is the Kronecker delta function and $b_{1/2}^m (x)$ is the Laplace coefficient:
\be
b_{1/2}^m(x) \equiv \frac{2}{\pi}\int_{0}^\pi \frac{\cos m\theta d\theta}{(1 - 2 x \cos \theta + x^2)^{1/2}}~.
\ee

For any given Fourier component $\Phi_m$ the disk response will have perturbations of the form $\delta \propto \exp(i m \phi - im \Omega_p t)$ giving
\ba
&-&i\tomega \frac{\Sigma}{c_s^2}\delta h + \frac{1}{r}\frac{\partial}{\partial r} (r \Sigma \delta u_r) + \frac{im}{r}\Sigma \delta u_\phi = 0,\\
&-&i \tomega \delta u_r - 2\Omega \delta u_\phi = -\frac{\partial}{\partial r}(\delta h + \Phi_m),\\
&-&i\tomega \delta u_\phi + \frac{\kappa^2}{2\Omega} \delta u_r = -\frac{im}{r}(\delta h + \Phi_m),
\ea
where $\kappa^2 \equiv \frac{2\Omega}{r}\frac{d}{dr}(r^2\Omega)$ is the square of the radial epicyclic frequency, $\tomega = m(\Omega_p - \Omega)$ is the Doppler shifted perturbation frequency, $\delta h \equiv \delta P/\Sigma$ is the enthalpy perturbation and $c_s^2 = \partial P/\partial \Sigma$ is the square of the sound speed.

Eliminating
$\delta u_r$ and $\delta u_\phi$ we find
\be
{\cal L}(\delta h + \Phi_m) + \frac{D}{c^2}\Phi_m = 0,  \label{secondorderdh}
\ee
where the operator ${\cal L}$ is defined as 
\ba
{\cal L} \equiv  \frac{d^2}{dr^2} + \left( \frac{d}{dr} \ln \frac{r\Sigma}{D} \right)\frac{d}{dr} &-& \frac{2m\Omega}{r\tomega} \left( \frac{d}{dr} \ln \frac{\Omega \Sigma}{D}\right)\nonumber \\
 &-&  \frac{m^2}{r^2} - \frac{D}{c^2}\, ,
\ea
and $D \equiv \kappa^2 - \tomega^2$. This is our basic working equation \citep[equivalent to equation (29) in][hereafter ZL06, with $n = 0$]{ZL06}. 

\subsection{The Differential Lindblad Torque}
Following \S 4 in ZL06 we can expand Eq. (\ref{secondorderdh}) in the vicinity of a Lindblad resonance $r_L$ where $D(r_L) = 0$, in the cold disk approximation. Changing variables to $x = (r-r_L)/r_L$ and keeping the important singular terms this gives
\be
\left(\frac{d^2}{dx^2} - \frac{1}{x} \frac{d}{dx} - \beta x\right) \delta h = \frac{1}{x}\Psi_m\, , \label{expandeq}
\ee 
where
\be
\Psi_m \equiv \left( r \frac{d}{dr} - \frac{2m\Omega}{\tomega}\right) \Phi_m \,,
\ee
and
\be
\beta \equiv \left. \frac{1}{c_s^2}\frac{dD}{d \ln r}  \right|_{r_L}\, .
\ee

To solve the above equation we define a new variable $\eta$ such that $\delta h = d\eta/dx$. Integrating and rearranging, Eq. (\ref{expandeq}) becomes
\be
\frac{d^2}{dx^2}\eta - \beta x \eta = -\Psi_m\, , \label{etaeq}
\ee
where we can absorb any integration constant into the definition of $\eta$. 

The homogeneous solutions for Eq. (\ref{etaeq}) are the Airy functions (Abramowitz \& Stegun 1964)
\ba
\eta_1 &=& {\rm Ai}(\beta^{1/3} x)\, ,  \\
\eta_2 &=& {\rm Bi}(\beta^{1/3} x)\, ,
\ea
while the general solution can be found using the method of variation of parameters
\ba
\eta = \frac{1}{W}\left(\eta_2 \int_0^x \eta_1 \Psi_mdx - \eta_1 \int_0^x \eta_2 \Psi_mdx\right)\nonumber \\
 + \hat{M}y_1 + \hat{N} y_2\, ,
\ea
where $W \equiv \eta_1 (d\eta_2/dx) - \eta_2 (d\eta_1/dx) = \beta^{1/3}/\pi$ is the Wronskian, and $M$ and $N$ are constants determined by the boundary conditions. 

If the function $\Psi_m$ does not vary strongly over the first Airy half-wavelength near the resonance then the solution can be approximated by
\ba
\eta &\simeq& \frac{\pi \Psi_m(r_L)}{\beta^{2/3}}\left[ \left( \int_0^\xi {\rm Ai}(\xi) d\xi + N \right){\rm Bi}(\xi)\right. \nonumber \\ 
&\qquad& \qquad \qquad \,\,-  \left. \left( \int_0^\xi {\rm Bi}(\xi) d\xi + M \right){\rm Ai}(\xi) \right]\, ,
\ea
where $\xi \equiv \beta^{1/3} x$, $M \equiv \hat{M}\beta^{2/3}/(\pi \Psi_m(r_L))$  and  $N \equiv \hat{N}\beta^{2/3}/(\pi \Psi_m(r_L))$ .

The asymptotic expansions of the Airy functions are well known. For $\xi < 0$ the behavior of the Airy functions is wavelike, while for $\xi > 0$ the ${\rm Ai}(\xi)$ behaves as a decaying exponential, while ${\rm Bi}(\xi)$ grows exponentially. Since ${\rm Bi}(\xi)$ grows exponentially and the physical solution must be bounded we require that as $\xi \rightarrow \infty$ the term multiplying ${\rm Bi}(\xi)$ must vanish. Hence we find that 
\be
N = -\int_0^\infty {\rm Ai}(\xi) d\xi = -\frac{1}{3}.
\ee

To determine $M$ we must consider the wave region in detail. 
The asymptotic expansions for ${\rm Ai}(\xi)$ and ${\rm Bi}(\xi)$ for $\xi \ll -1$ are (taken from ZL06)
\ba
{\rm Ai}(\xi) &\sim& \pi^{-1/2} (-\xi)^{-1/4} \sin X(\xi)\, ,\\
\int_0^\xi {\rm Ai}(\xi)&\sim& -\frac{2}{3} + \pi^{-1/2} (-\xi)^{-1/4} \cos X(\xi)\, ,\\
{\rm Bi}(\xi) &\sim& \pi^{-1/2} (-\xi)^{-1/4} \cos X(\xi)\, ,\\
\int_0^\xi {\rm Bi}(\xi)&\sim& -\pi^{-1/2} (-\xi)^{-1/4} \sin X(\xi)\, ,
\ea
where 
\be
X(\xi) = \frac{2}{3}(-\xi)^{3/2} + \frac{\pi}{4}\, .
\ee
This gives the asymptotic solution 
\ba
\delta h \sim \frac{i \pi^{1/2}\Psi_m}{2\beta^{1/3}}(-\xi)^{1/4}[ (&1&-iM) e^{iX(\xi)}\nonumber
\\ &+& (1 + iM) e^{-iX(\xi)}]\, .\label{Meq}
\ea

As in ZL06 we have for the outer Lindblad resonance ($r_{\rm OLR}$) an outgoing wave if the solution has the form $\exp(+iX(\xi))$. This outgoing wave solution is achieved by setting $M = i$, and yields
\be
\delta h(r > r_{\rm OLR}) = \frac{i\pi \Psi_m}{\beta^{1/3}}(-\xi)^{1/4} e^{iX(\xi)}\, .
\ee

For the inner Lindblad resonance ($r_{\rm ILR}$) an ingoing wave (propagating away from the resonance) has the form $\exp(-iX(\xi))$ (since the group velocity and phase velocity have opposite sign inside the corotation resonance), and an outgoing wave (propagating towards the resonance) has the form $\exp(+iX(\xi))$. With no reflecting inner boundary in the disk the ingoing wave correctly satisfies the boundary condition (no waves are reflected back) and the correct solution is obtained by setting $M = -i$ such that 
\be
\delta h(r < r_{\rm ILR}) = -\frac{i\pi \Psi_m}{\beta^{1/3}}(-\xi)^{1/4} e^{-iX(\xi)}\, . 
\ee

The torque can be calculated by determining the angular momentum flux \citep[see GT79;][]{LBK72, Tanaka02}
\be
F(r) \simeq \frac{\pi m r \Sigma}{D}{\rm Im}\left( \delta h \frac{d\delta h^*}{dr}\right) \label{flux}
\ee
where $^*$ denotes the complex conjugate. The torque on the disk due to each resonance is then given by the flux evaluated in the wave region away from the resonance.
This gives 
\be
T_m(r_{\rm OLR}) = F(r > r_{\rm OLR}) = \left. \frac{m\pi^2\Sigma\Psi_m^2}{r dD/dr}\right|_{r_{\rm OLR}} \label{TOLReq}
\ee
for the outer Linbdlad resonance and
\be
T_m(r_{\rm ILR}) = F(r < r_{\rm ILR}) = \left. -\frac{m\pi^2\Sigma\Psi_m^2}{r dD/dr}\right|_{r_{\rm ILR}}\label{TILReq}
\ee
for the inner Lindblad resonance, which agrees with W86 and GT79. 

As discussed in W97 and \citet{Artymowicz93, Artymowicz94} the torque cutoff/pressure buffer effect can be included if $\Psi_m$ is multiplied by a factor of $[1 + 4(mc/(r\kappa))^2]^{-1/2}$, the $D$ in \eqref{TOLReq} and \eqref{TILReq} is replaced by $D' \equiv \kappa^2 - \tomega^2 + m^2c_s^2/r^2$ (to account for the excess pressure due to azimuthal winding), and the torque expressions are instead evaluated at the modified inner and outer Lindblad resonances where $D' = 0$. 

\begin{figure}
\epsfig{file=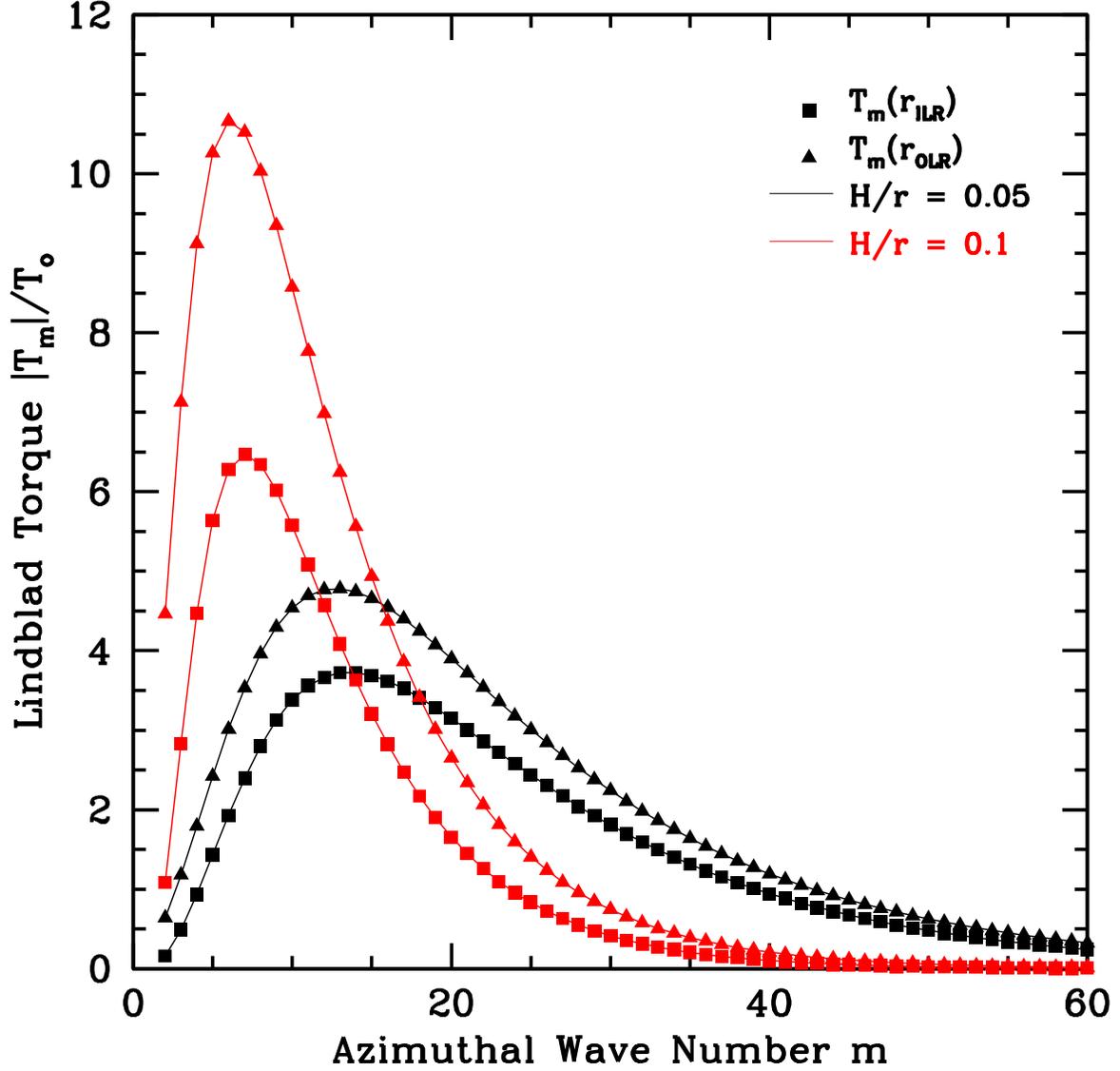, width=\linewidth}
\caption{The normal inner and outer Lindblad torques for various values of the disk thickness $H/r$ evaluated including the effect of the pressure buffer (see W97), and normalized by the value $T_o \equiv (M_p/M_*)^2 \Sigma r^4 \Omega^2 (H/r)^{-3}$ evaluated at the planet location, where $M_*$ is the mass of the host star. The torques are cutoff for $m > \Delta m_{\rm cutoff} \propto (H/r)^{-1}$.}
\label{Difffig}
\end{figure}

Evaluating the inner and outer Lindblad torques we see in Figure \ref{Difffig} that for waves traveling away from resonances the outer torque will be stronger than the inner torque and the net migration must therefore be inwards in Type I migration. 

\subsection{Torque Due to a Trapped Wave}

We can generalize the above results \citep[see][for a more formal discussion for the homogeneous solution]{Tsang08}  for waves traveling far from the resonance by defining
\be
k \simeq \sqrt{D/c_s^2} 
\ee
and substituting
\be
X(\xi) = \pm \int_{r_L}^{r} k dr  + \frac{\pi}{4}, \qquad(-\xi)^{1/2} = \mp r_L k/\beta^{1/3} \, ,
\ee
where the signs are for the outer and inner Lindblad resonances respectively. 

If there exists a boundary at some position $r_{\rm in} < r_{\rm ILR}$ with a complex reflection coefficient ${\cal R}$ then
\be
\delta h(r_{\rm in}) \propto \exp\left( i\int_{r_{\rm in}}^{r} k~dr\right) + {\cal R}\exp \left( -i\int_{r_{\rm in}}^{r} k ~dr \right)\, .\label{bcond}
\ee

To satisfy Eq. (\ref{bcond}) and (\ref{Meq}) we must then have
\be
\frac{1-iM}{1+iM}  = {\cal R} e^{i2\Theta}\, ,
\ee
where 
\be
\Theta \equiv \int_{r_{\rm in}}^{r_{\rm IL}} k dr + \pi/4\, .
\ee
This gives us
\ba
M &=& i \frac{e^{[i\Theta + (\ln {\cal R})/2]} -  e^{-[i\Theta + (\ln {\cal R})/2]}}{e^{[i\Theta + (\ln {\cal R})/2]} +  e^{-[i\Theta + (\ln {\cal R})/2]}} \nonumber \\
 &=& -\tan[\Theta - i(\ln {\cal R})/2]\, .
\ea

Applying Eq (\ref{Meq}) to Eq. (\ref{flux}) we find
\ba
T_m(r_{\rm ILR}) &=& F(r < r_{\rm ILR})\nonumber \\
 &=& -m\pi^2(iM)\left(\frac{\Sigma\Psi_m^2}{r dD/dr}\right)_{r_{\rm ILR}} \, .
\ea

When ${\cal R} = 0$ we have $M = -i$ recovering the no-reflection case. However for non-zero ${\cal R}$ we have
\be
M = -\tan[\Theta - {\rm Arg}({\cal R}^{-1/2}) + i \ln|{\cal R}|^{-1/2}]~.
\ee
When $\Theta - {\rm Arg}({\cal R}^{-1/2}) = \pi(n + 1/2)$ and the waves launched at the Lindblad resonance becomes trapped then we have $M = -\frac{i}{\tanh[(\ln|R|^{-1/2})]}$
which gives the torque
\be
T_m =  -\left( \frac{m\pi^2\Sigma\Psi_m^2}{r dD/dr}\right)_{r_{\rm ILR}}\left(\frac{1}{\tanh[\ln |{\cal R}|^{-1/2}]}\right) \, .
\ee
As $|{\cal R}| \rightarrow 1$ (e.g. for the free boundary condition $\delta h = 0$ we have $R = e^{i\pi}$) the torque due to the resonant wave approaches $T_m \rightarrow \infty$. However, we must also take into account the damping of the propagating wave. 

\subsection{Wave Damping}
If the damping is too strong, such that the damping length is comparable to the first Airy half-wavelength, then the torque cannot be easily related to the flux as in Eq. (\ref{flux}). However, for damping length much greater than the driving length we may continue to use the flux evaluated near the Lindblad resonance as a measure of the net torque exerted on the disk. 

Introducing a small damping term into the system will contribute an imaginary component to the wavenumber, $k = k_r + i k_i$, such that
\be
\Theta = \Theta_r + i \Theta_i \equiv \int_{r_{\rm in}}^{r_{\rm IL}} k_r + i k_i dr \, ,
\ee 
which gives the result
\be
T_m = \left( \frac{m\pi^2\Sigma\Psi_m^2}{r dD/dr}\right)_{r_{\rm ILR}} i \tan[\Theta_r + i\Theta_i +i\ln{\cal R}^{-1/2}]~.
\ee
When $\Theta_r + {\rm Arg}({\cal R}^{-1/2}) = \pi(n + 1/2)$ and the density wave is trapped between the inner boundary and the inner Lindblad resonance we have the torque 
\be
T_m =  -\left( \frac{m\pi^2\Sigma\Psi_m^2}{r dD/dr}\right)_{r_{\rm ILR}} \left(\frac{1}{\tanh[\ln |{\cal R}|^{-1/2} + \Theta_i]} \right)~.
\ee

We define the resonant gain $\Gamma_m$ as
\be
\Gamma_m  \equiv \frac{T_m}{(T_m)_{{\cal R}\rightarrow 0}} = \frac{1}{\tanh[\ln |{\cal R}|^{-1/2} + \Theta_i]}~,
\ee
the gain in the inner Lindblad torque due to the wave being resonant in the disk.

Here, for simplicity, we take the viscosity to be independent of the perturbed quantities so that viscous damping takes the form of given by W97
\be
\Theta_i = \frac{|(r-r_{L})/H|^{3/2}}{\Delta} \, ,
\ee
where $\Delta \simeq \alpha^{-1} m^{-1/2} (H/r)$, $H = c_s/\Omega$ is the disk scale height, and $\alpha$ is the \citet{Shakura73} viscosity parameter.

For strong wave damping $\Theta_i$ is large and $\Gamma_m \rightarrow 1$, the normal Lindblad torque. However, for weak damping $\Theta_i \rightarrow 0$ with perfect reflection $|{\cal R}| \rightarrow 1$ we have $\Gamma_m \rightarrow \infty$, allowing for very large inner Lindblad torque..

\section{Halting Type I Migration: Linear Case}

\begin{figure}
\epsfig{file=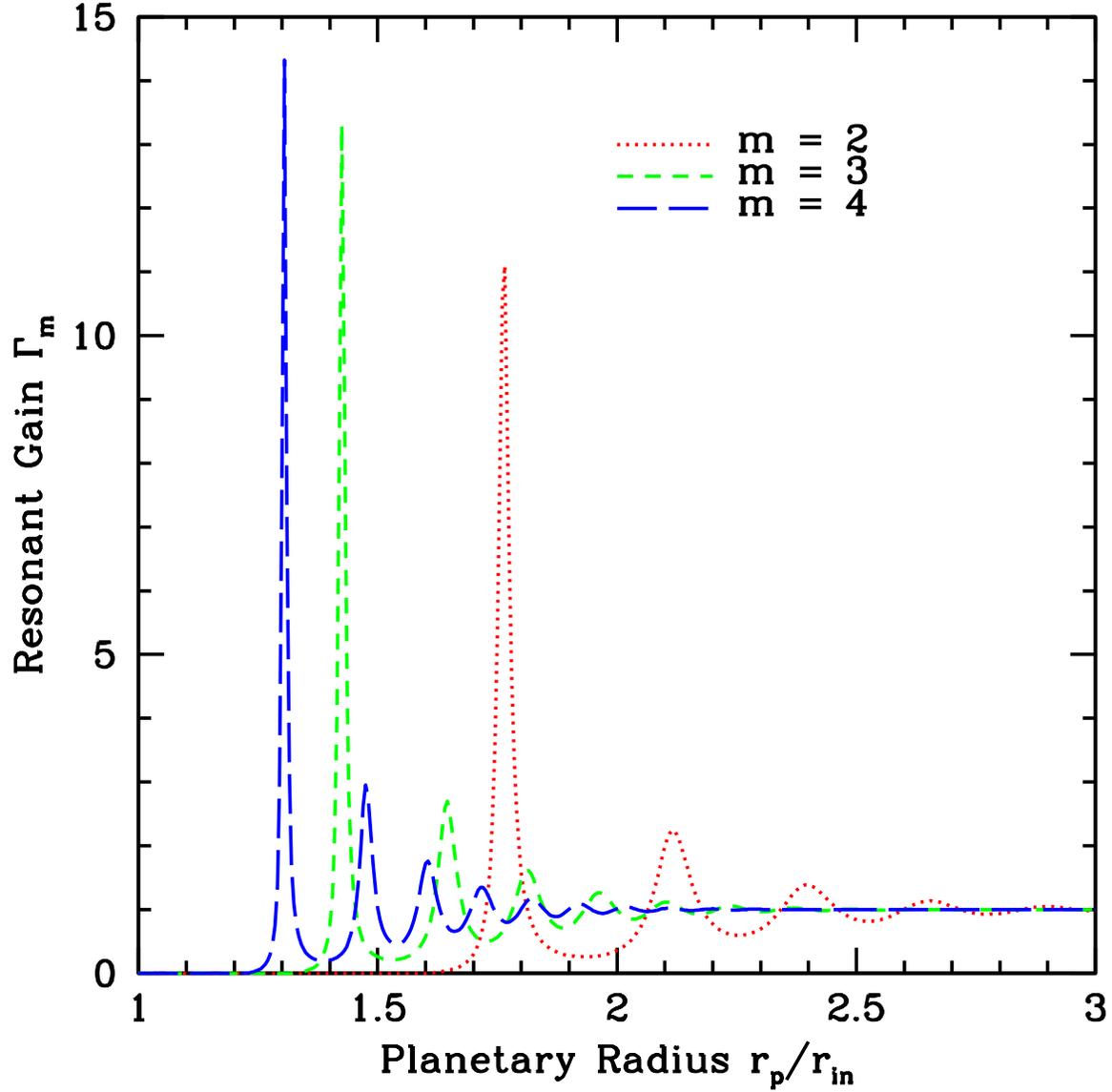, width=\linewidth}
\caption{Resonant gain $\Gamma_m$ for the first three azimuthal modes for linear viscous damping with $H/r = 0.05$ and $\alpha = 10^{-3}$. Far from the disk edge $\Gamma_m \rightarrow 1$. The gain increases strongly for waves that are resonant in the disk.}
\label{Gainfig}
\end{figure}

For the cases where the corotation torque is negligible, Type I migration will be halted if the total torque due to the inner Lindblad resonances exceed the torques due to the outer Lindblad resonances. The resonant gain as a function of $r_p/r_{\rm in}$ are shown for the first few $m$ in Figure \ref{Gainfig}. 


There exists a torque cutoff (GT80, W97) such that only a finite range of $m$ contribute strongly to the total torque (see Figure \ref{Difffig}). As the planet migrates through the disk it sweeps through various resonances for the global modes. If the excess torque generated by a $m$-mode that is in resonance exceeds the torque deficit due to all the other $m$ within the torque cutoff then the migration can be stopped by the trapped wave resonance.

We can estimate the maximum viscosity parameter $\alpha$ such that the migration can be halted by this effect by setting the maximal gain $\Gamma_m$  roughly equal to the width of the torque cutoff
\be
\Delta m_{\rm cutoff} \equiv f \left( \frac{r}{H}\right) \, ,
\ee
where $f$ is a factor of few that absorbs the variation of the torque for azimuthal modes within the torque cutoff. This is the worst case assumption, applicable since away from the strong resonant peak where the gain is $\Gamma_{res}= 1/\tanh(\Theta_i + \ln|{\cal R}|^{1/2})$, we have a reduced gain, $\sim 1/\Gamma_{res} \ll 1$. Thus the azimuthal mode in resonance must make up for, at most, the rest of the modes within the cutoff.  We then have
\be
\frac{1}{\tanh(\Theta_i + \ln|{\cal R}|^{-1/2})} \go \Delta m_{\rm cutoff} \, .
\ee
If we take the boundary condition to be $\delta h(r_{\rm in}) = 0$ then we have $|{\cal R}| = 1$, and
\be
\Theta_i \lo \tanh^{-1}\left[\frac{1}{f} \left( \frac{H}{r}\right)\right]\simeq \frac{1}{f}\left(\frac{H}{r} \right) \, ,
\ee
where the last approximation holds for small $H/r$. The above condition says that Type I migration can be halted, if the number of e-foldings of wave amplitude lost across the resonant cavity is smaller than $1/\Delta m_{\rm cutoff}$.

\begin{figure}[t!]
\epsfig{file=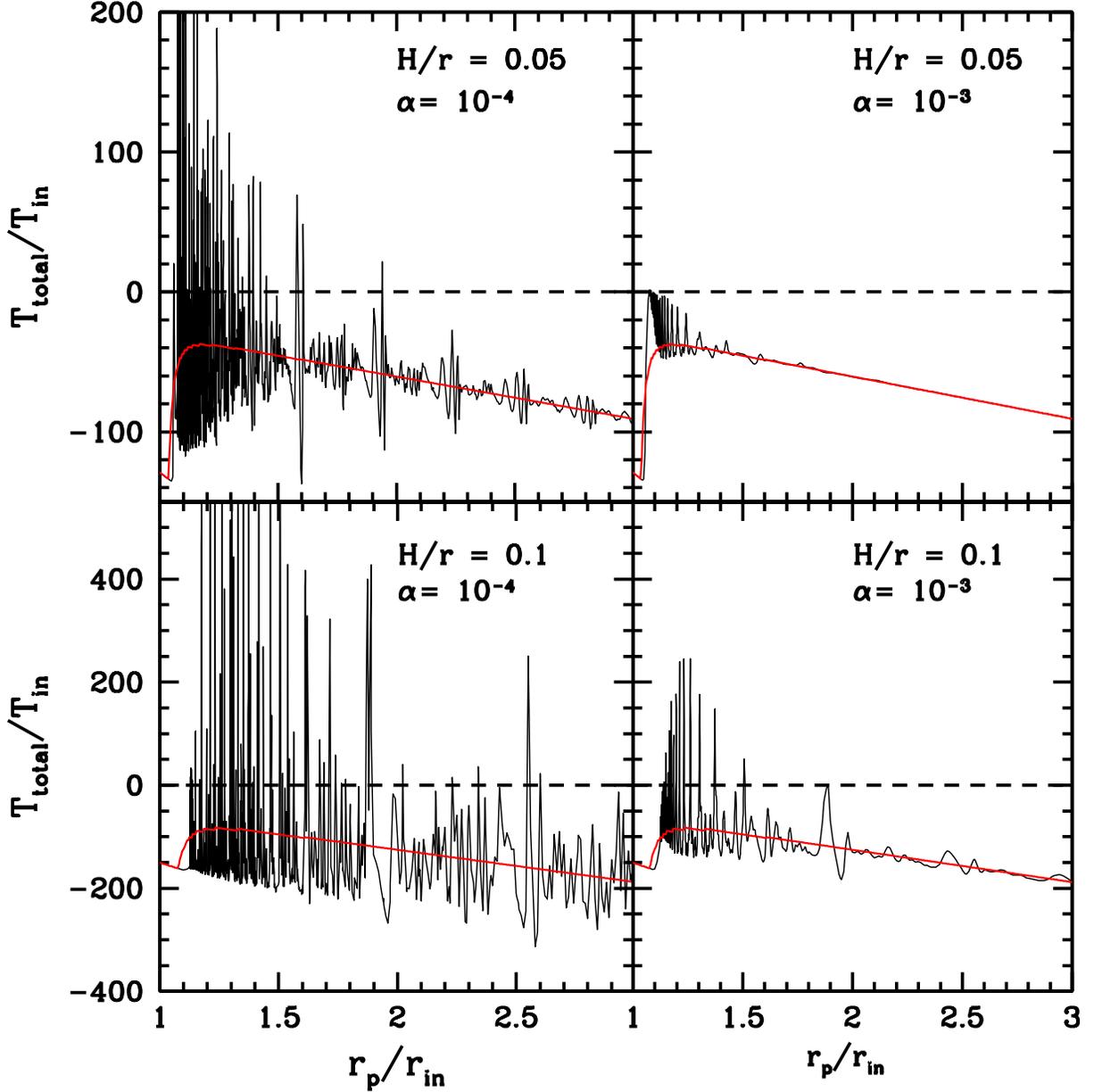, width=\linewidth}
\caption{Net Lindblad torque, $T_{\rm total}$, for the $m\leq 60$ modes as a function of planet orbital radius $r_p$. The torque is negative except at the resonant peaks where the gain due to wave trapping is high enough for the inner Lindblad torque to overcome the outer Lindblad torque. The net Lindblad torque is normalized by $T_{\rm in} \equiv (M_p/M_*)^2 \Sigma r^4 \Omega^2 (H/r)^{-3}$ evaluated at the inner edge of the disk $r = r_{\rm in}$. The red lines indicate the differential Lindblad torque without considering disk resonance, while the black shows the differential Lindblad torque including the effect of disk resonance. When corotation torque is negligible planets can become trapped where the net Lindblad torque crosses zero with negative slope.}
\label{torquefig}
\end{figure}

This yields, for viscous dissipation,
\be
\alpha \lo \frac{1}{f}\frac{1}{m^{1/2}} \left( \frac{H}{r}\right)^2 \left(\frac{r_{\rm ILR} - r_{\rm in}}{H}\right)^{-3/2}~.
\ee
If we assume that $r_{\rm in}$ is located at the minimum distance for a resonance, the location of the first node, then we have the wavelength condition
\be
\int_{r_{\rm in}}^{r_{\rm ILR}} \sqrt{\frac{D}{c_s^2}} dr + \pi/4  \simeq \pi/2 \, .
\ee
Near the Lindblad resonance we have $D \simeq  (r-r_{\rm ILR})(dD/dr)_{r_{\rm ILR}}$ giving
\be
\left(\frac{r_{\rm ILR}-r_{\rm in}}{H}\right)^{3/2} \simeq \frac{3\pi}{8} \left[3 m \left(\frac{r_{\rm ILR}}{r_p}\right)^{1/2}\frac{H}{r_p} \right]^{-1/2}\, .
\ee
Thus if the viscosity parameter $\alpha$ is smaller than
\be
\alpha \lo \frac{1}{f} \frac{8}{\sqrt{3}\pi}\left(\frac{r_{\rm ILR}}{r_p}\right)^{1/4}\left(\frac{H}{r} \right)^{5/2} \sim \left( \frac{H}{r}\right)^{5/2} \label{alphaeq}
\ee
type I migration can be halted by trapped modes in the disk.

The net Lindblad torque for the first $m \leq 60$ modes are shown in Figure \ref{torquefig} for various combinations of disk scale height $H$ and viscosity $\alpha$. As a planet migrates slowly inwards from large $r$, the net Lindblad torque, $T_{\rm total}$, is negative, driving the planet inwards. When the planet encounters a disk resonance strong enough such that $T_{\rm total} > 0$ the planet can become trapped. Further migration inwards towards the resonance causes the net Lindblad torque to become positive, driving the planet outwards, while outwards migration away from the resonance leads to a net negative Lindblad torque, driving the planet inwards, effectively trapping the planet where $T_{\rm total} \simeq 0$.

In Figure \ref{torquefig}, for low viscosity ($\alpha = 10^{-4}$), we see many locations of effective planet trapping for disks with both $H/r = 0.1$ and $0.05$. However for higher viscosity ($\alpha = 10^{-3}$) we only see trapping for $H/r = 0.1$, consistent with our estimate \eqref{alphaeq}. 

Including the effect of the corotation torque $T_{\rm CR}$ requires a location where $\sum T = T_{\rm CR} + T_{\rm ILR} + T_{\rm OLR} = 0$ to trap the planet. The value of the unsaturated corotation torque in an isothermal disk is proportional to the vortensity gradient at the corotation point (GT79), 
\be
T_{\rm CR} \propto - \left.\frac{d\ln [\kappa^2/(2\Sigma \Omega)]}{d\ln r}\right|_{r_{\rm CR}}\,,
\ee
and can be larger than the non-resonant net Lindblad torque near sharp density transitions. \citet{Baruteau08} also showed that the corotation can contribute a significant torque in the presence of entropy gradients in non-barotropic disks. Depending on the sign of the density and entropy gradients in the disk at the planet location the corotation torque can make it easier or more difficult to trap a planet using the trapped wave disk resonance discussed above.

\section{Nonlinear Resonance}
\citet{GR01} and \citet{Rafikov02} showed that the spiral density waves generated by a perturbing planet typically steepen as they propagate and can quickly act to shock and damp out the wave. For waves resonant in the disk this would lead to a non-linear resonance, in which the primary means of dissipation involves wave shocks. Detailed examination of the non-linear response of the disk is left for later work, and may perhaps be best approached numerically. However, we note that typically for forced non-linear resonators, such as shock tubes, the resonator quality factor, $Q$, effectively our resonant gain, scales as the inverse square root of the forced amplitude \citep{Rudenko09}. This implies that the non-linear resonant gain decreases with with an increase in the planet mass, and that there is a cutoff mass above which the resonance can no longer halt Type I migration. 

\section{Discussion}
We have shown that for small planets the linear response of a disk resonance caused by an inner reflecting edge can act to trap planets at a few times the inner disk radius. The exact location of this trapping depends on the details of the disk parameters and dissipation mechanism. For larger planets, the disk will likely undergo a non-linear resonance, which typically has non-linear resonant gain decreasing with increasing planet mass. This implies that above a certain cutoff mass, the resonant gain due to the non-linear resonance will be insufficient to halt migration, and the planets will continue to migrate inwards through the inner disk edge. There they will be halted at $r_{\rm halt} $ when the lowest order outer Lindblad resonance is located within the inner edge of the disk. The corotation torque is also important near these edges, but is susceptible to saturation, unless the viscosity is sufficiently high. If unsaturated the corotation torque is mainly active when the planet is very close to the disk edge and the density gradient (and thus vortensity gradient) are strongest. However, further from the disk edge where the wave trapping resonance is active, both effects may be important in determining the net torque.

This naturally generates two populations of planets, with halting radius a few times the inner disk edge for smaller planets, and with larger planets halted inside the inner disk edge. This process may help to explain the observations of planets close to solar type stars by \citet{Howard11} which showed that larger planets were halted with typical periods of $\sim 2$ days, corresponding to $r_{\rm halt} \sim 0.03$ AU, while smaller planets were halted with typical periods of $\sim 7$ days, $r_{\rm halt} \sim 0.07$ AU, consistent with inner disk edges with periods of  $\sim 4$ days corresponding to $r_{\rm in} \sim 0.05$ AU.

This process may also be important for planets located near other reflecting edges within the disk, such as the edges of magnetic dead zones \citep[see e.g.][]{Gammie96}, or the edge of a gap in the disk opened by a giant planet \citep[see e.g.][]{Lin79}. 

The effect of turbulent density fluctuations on this wave resonance may be to ratchet the system through successive ``traps'', if the variation is on a scale similar to the separation between the Lindblad resonances, such that the relative amplitudes of the Lindblad torques are affected. Larger resonant peaks will be more difficult to overcome, however a detailed study of the effect of turbulence is beyond the scope of the current work.

\section*{Acknowledgements}

I would like to acknowledge kind advice and useful conversations regarding this work with Peter Goldreich, Dong Lai, Christian Ott, John Johnson, Scott Gregory, and Jason Wright.  This work was carried out with the support of a Sherman Fairchild Postdoctoral Fellowship at the California Institute of Technology.

\renewcommand{\bibsection}{\section{References}} 

\end{document}